# Leptogenesis in the Symmetric Phase of the Early Universe: Baryon Asymmetry and Hypermagnetic Helicity Evolution

**V. B. Semikoz[a,*] and A. Yu. Smirnov[a,b]**

[a] *Pushkov Institute of Terrestrial Magnetism, Ionosphere, and Radio Wave Propagation, Russian Academy of Sciences (IZMIRAN), Troitsk, Moscow oblast, 142190 Russia*

[b] *National University of Science and Technology "MISIS," Moscow, 119991 Russia*

**e-mail: semikoz@yandex.ru*

**Abstract**—We investigate the evolution of the baryon asymmetry of the Universe (BAU) in its symmetric phase before the electroweak phase transition (EWPT) induced by leptogenesis in the hypermagnetic field of an arbitrary structure and with a maximum hypermagnetic helicity density. The novelty of this work is that the BAU has been calculated for a continuous hypermagnetic helicity spectrum. The observed BAU $B_{\text{obs}} = 10^{-10}$ that can be in large-scale hypermagnetic fields satisfying the wave number inequality $k \le k_{\max}$ grows with increasing $k_{\max}$. We will also show that the initial right-handed electron asymmetry $\xi_{eR}(\eta_0)$ used in our leptogenesis model as a free parameter cannot take too large values, $\xi_{eR}(\eta_0) = 10^{-4}$, because this leads to a negative BAU by the EWPT time. In contrast, a sufficiently small initial right-handed electron asymmetry, $\xi_{eR}(\eta_0)$, provides its further growth and the corresponding BAU growth from zero to some positive value, including the observed $B_{\text{obs}} = 10^{-10}$.

## 1. INTRODUCTION

Astrophysical magnetic fields affect the cosmic-ray propagation, the stellar (solar) activity, etc., while their origin is still an open question in astrophysics and cosmology [1–3]. Since Maxwell's equations are linear in fields **E** and **B**, there must be some seed (magnetic) field needed to switch on a dynamo leading to field amplification to the observed galactic magnetic field strengths, $B_{\text{gal}} \sim 10^{-6}$ G. There are two possibilities of searching for such a seed in a galaxy: (i) an astrophysical one, for example, in scenarios that take into account supernova explosions with the ejection of a magnetohydrodynamic (MHD) plasma with a frozen-in magnetic field into intergalactic space and (ii) a cosmological scenario that envisages the existence of seed fields (and considers their evolution) over the radiation-dominated and dust-like stages of the early Universe. In this paper, we rely on the second scenario (ii). An upper limit on the cosmological magnetic field (CMF), $B < 10^{-10}$–$10^{-9}$ G, has long been known, for example, from observations of the Faraday rotation of the radio-emission polarization plane [4]. The first evidence for the presence of CMF in the intergalactic medium that can survive up to the present epoch is related to the prediction of a lower limit for the CMF amplitudes, $B_{\text{CMF}} > 10^{-16}$–$10^{-14}$ G, that follows from satellite observations of high-energy photons (in particular, from the Fermi experiment) [5, 6], which is a new confirmation of the CMF concept used here.

In this paper, we will be interested in the fundamental problem of baryon asymmetry generation in the primordial CMF existing before the electroweak phase transition (EWPT) in the early Universe. To clarify the nature of this field, we will note that the Maxwellian field $A_\mu$ is the trace of the Abelian $U(1)_Y$ hypercharge field $Y_\mu$. The latter exists in the primordial plasma before the EWPT as the only massless (long-range) gauge field, as distinct from the non-Abelian components $W^3_\mu$ possessing a "magnetic" mass gap in the plasma $\sim g^2T$, i.e., vanishing on large scales. Both fields enter into the canonical relation $A_\mu = \cos\theta_W Y_\mu + \sin\theta_W W^3_\mu$, where $\sin^2\theta_W = 0.23$ is the Weinberg parameter in the Standard Model (SM). This difference in spatial scales explains why the boundary condition $A_\mu = \cos\theta_W Y_\mu$ at the EWPT time $t = t_{EW}$ near the boundary of the new-phase bubble[1] should be used for a massless photon.

Thus, the hypermagnetic field (HMF) $\mathbf{B}_Y = \nabla \times \mathbf{Y}$ formed before the EWPT and its helicity density $h_Y = \mathbf{Y} \cdot \mathbf{B}_Y$ turn out to be important sources for such characteristics of the Maxwellian field as its initial strength $B$, initial correlation length $\Lambda$, and initial magnetic helicity density $h = \mathbf{A} \cdot \mathbf{B}$. There are other important problems related to the change in HMF helicity density $dh_Y/dt = -2\mathbf{E}_Y \cdot \mathbf{B}_Y$. At the one-loop level, $dh_Y/dt$ is proportional to the fermion number violation $\partial_\mu j^\mu \sim \mathbf{E}_Y \cdot \mathbf{B}_Y \neq 0$ owing to the Abelian anomaly or the Fermi number "sits" in the HMF [7].

[1] We assume here a first-order EWPT maintained by a strong hypermagnetic field.

The problem of lepton asymmetry evolution via the Abelian anomaly in a helical HMF is directly related to the growth the baryon asymmetry of the Universe (BAU). Note that the process of leptogenesis itself in a HMF has already been studied in our recent paper [8], and this study was performed for an arbitrary HMF configuration with a maximum helicity density. In our previous papers [9, 10], we also considered the BAU evolution based on specific one-dimensional HMF configurations: a Chern–Simons wave $Y_0 = 0$, $\mathbf{Y} = Y(t)(\sin k_0 z, \cos k_0 z, 0)$ with a fixed wave number $k_0 =$ const and a maximum helicity density. Thus, we ignored the inverse cascade that is needed in the case of a more realistic continuous helicity spectrum for an arbitrary $3D$ HMF configuration.

The main goal of this paper is a complete description of the BAU growth in a helical HMF for an arbitrary $3D$ HMF configuration up to the EWPT time. We have to take into account the sphaleron vacuum–vacuum transitions in our model when the left-handed lepton asymmetry in an equilibrium plasma is taken into account. This reduces the number of produced left-handed leptons and, accordingly, the BAU owing to the global charge conservation law $B/3 - L_e =$ const in an external HMF.

Our scenario is as follows. We consider the plasma of the hot Universe before the EWPT at the stage $T_{RL} > T > T_{EW}$, when the left-handed leptons $L = (\nu_{eL} e_L)^T$ come into equilibrium with the primordial right-handed electrons $e_R$ through the Higgs inverse decay $e_R \bar{e}_L \longrightarrow \varphi^{(0)}$, $e_R \nu_{eL} \longrightarrow \varphi^{(-)}$.[2] This occurs as the Universe cools down to a temperature $T_{RL} \sim 10$ TeV, when the Higgs decay rate $\Gamma_{RL} \sim T$ becomes larger than the Hubble expansion rate of the Universe $H \sim T^2$, $\Gamma_{RL} \geq H$.

It was shown in [12] that a seed HMF gives rise to a Chern–Simons contribution to the effective SM Lagrangian of the field $Y_\mu$ through the polarization effect that is related to the nonzero mean (macroscopic) pseudo-vector lepton current $j_{i5} = \langle \bar{\psi} \gamma_i \gamma_5 \psi \rangle \sim B_i^Y \neq 0$. The appearance of left-handed fermions leads to an additional polarization effect due to the macroscopic currents of left-handed leptons in the seed HMF $\mathbf{B}_Y$,

$$J_{i5}^{(e)} = \langle \bar{\psi}_{eL} \gamma_i \gamma_5 \psi_{eL} \rangle \sim \mu_{eL} B_i^Y,$$

$$J_{i5}^{(\nu)} = \langle \bar{\nu}_{eL} \gamma_i \gamma_5 \nu_{eL} \rangle \sim \mu_{eL} B_i^Y,$$

where the left-handed lepton chemical potential $\mu_{eL}$ for the doublet $L = (\nu_{eL} e_L)^T$ coincides with the left-handed neutrino chemical potential, $\mu_{eL} = \mu_{\nu_{eL}}$.

Given the evolution of the left-handed lepton asymmetry $(n_{eL} - n_{\bar{e}L}) \sim \mu_{eL}(t)$ owing to the Abelian anomaly at temperatures $T_{EW} < T < T_{RL}$ and given the interaction of left-handed fermions with sphalerons, we also extend the scenario [7, 13] based on leptogenesis due to the evolution of the asymmetry of the right-handed electrons alone $(n_{eR} - n_{\bar{e}R}) \sim \mu_{eR}(t) \neq 0$ in the same HMFs $\mathbf{B}_Y \neq 0$.

Below in Section 2, we derive a kinetic equation for the density of the HMF spectrum in Fourier representation using conformal variables. Such a spectrum depends on the lepton asymmetries that develop in a self-consistent HMF, as described in Section 2.2. Then in the main Section 3, we calculate the BAU by using the (t' Hooft's) conservation law $B/3 - L_e =$ const and numerically solving the self-consistent nonlinear kinetic equations for the lepton number $L_e$ and a continuous helicity density spectrum $\tilde{h}_Y$. In Section 4, we discuss our results by comparing them with some of the previous BAU calculations (for a monochromatic helicity density spectrum) in the same leptogenesis scenario.

## 2. LEPTOGENESIS IN HYPERMAGNETIC FIELDS

In the Standard Model $U(1)_Y$, the Abelian anomalies arising in a hypercharge field $Y_\mu$,

$$\frac{\partial j^\mu_{R,L}}{\partial x^\mu} = \pm \frac{g'^2 Y_{R,L}^2}{64\pi^2} Y_{\mu\nu} \tilde{Y}^{\mu\nu}, \quad (1)$$

violate the conservation law for the corresponding lepton numbers.

Here, $Y_R = -2$ and $Y_L = -1$ are the hypercharges of the right- and left-handed leptons, respectively, $Y_{\mu\nu}$ and dual $\tilde{Y}_{\mu\nu}$ are the hypercharge field strengths, $g' = e/\cos\theta_W$ is the gauge coupling constant in the SM. The upper (lower) sign on the right-hand side of (1) corresponds to the right-handed (left-handed) currents, $j_R^\mu = \bar{\Psi}_R \gamma^\mu \Psi_R$ and $j_L^\mu = \bar{\Psi}_L \gamma^\mu \Psi_L$, where $\Psi_R = (1 + \gamma_5)\Psi/2$ and $\Psi_L = (1 - \gamma_5)\Psi/2$ are the right- and left-handed bispinor fields, respectively.

[2] It should be noted that the process of Higgs boson decays (inverse decays) is not the only reaction channel leading to a lepton chirality flip. For example, the scattering, in particular, of $e_R$ by a Higgs boson $e_R H \longleftrightarrow L_e A$, where $A = Y$ or $W$ are the gauge fields, can be such a process [11]. Evaluating the role of left-handed leptons (electrons) in baryogenesis at least for one of the reaction channels as an example is important to us. One of the authors (V. S.) thanks Kimmo Kainulainen for the comment on this theme when discussing the series of our preceding works.

*2.1. Hypermagnetic Helicity before the EWPT*

If the medium is at rest as a whole, then Faraday's equation describing a HMF $\mathbf{B}_Y = \nabla \times \mathbf{Y}$ reads[3]

$$\frac{\partial \mathbf{B}_Y}{\partial t} = \nabla \times \alpha_Y \mathbf{B}_Y + \eta_Y \nabla^2 \mathbf{B}_Y, \tag{2}$$

where the hypermagnetic helicity coefficient $\alpha_Y$ at temperatures $T_{RL} > T > T_{EW}$ is calculated from the left- and right-handed electron chemical potentials $\mu_{eR}$ and $\mu_{eL}$ [9, 10],

$$\alpha_Y(T) = \frac{g'^2(\mu_{eR} + \mu_{eL}/2)}{4\pi^2 \sigma_{\text{cond}}}, \tag{3}$$

and $\eta_Y = (\sigma_{\text{cond}})^{-1}$ is the hypermagnetic diffusion coefficient, and $\sigma_{\text{cond}}(T) \approx 100T$ is the hot plasma conductivity. We emphasize that the $\alpha_Y$ effect in Faraday's equation (2) arises from the addition of the Chern–Simons term $L_{CS} \sim \mathbf{Y} \cdot \mathbf{B}_Y$ to the effective Lagrangian of the hypercharge field interacting with particles in the SM plasma and attributable to the polarization effect induced by the HMF [12]. Multiplying Eq. (2) by the corresponding vector potential and adding the analogous construction obtained by multiplying the evolution equation for the vector potential by the field, after integration over space we get the evolution equation for the hypermagnetic helicity $H_Y = \int d^3x \mathbf{Y} \cdot \mathbf{B}_Y$

$$\begin{aligned} \frac{dH_Y}{dt} &= -2\int_V (\mathbf{E}_Y \cdot \mathbf{B}_Y) d^3x \\ -\oint [Y_0 \mathbf{B}_Y + \mathbf{E}_Y \times \mathbf{Y}] d^2S &= -2\eta_Y(t) \int d^3x (\nabla \times \mathbf{B}_Y) \cdot \mathbf{B}_Y \\ &+ 2\alpha_Y(t) \int d^3x B_Y^2(t). \end{aligned} \tag{4}$$

For the single symmetric phase before the EWPT, we omit the surface integral $\oint(\ldots)$ in the last row in (4), because the hypercharge fields vanish at infinity. However, such a surface integral can be important at the interface between different phases during the EWPT, $T \sim T_{EW}$. The authors in [14] study how the hypermagnetic helicity flux penetrates through the surface separating the symmetric phase and the phase with broken symmetry and how the hypermagnetic helicity density $h_Y = \mathbf{B}_Y \cdot \mathbf{Y}$ transforms into the magnetic helicity density $h = \mathbf{B} \cdot \mathbf{A}$ at the first-order EWPT time.

[3] We neglected the change in bulk velocity in the plasma described by the Navier–Stokes equation everywhere in the text, because the length scale of the velocity change $\lambda_V$ is much smaller than the HMF correlation length, $\lambda_V \ll k^{-1}$; in other words, the infrared modes of the HMF are virtually independent of the plasma velocity. In addition, the bulk velocity $\mathbf{v}$ does not contribute to the helicity evolution $dh_Y/dt \sim (\mathbf{E}_Y \cdot \mathbf{B}_Y)$ when the generalized Ohm law is used, $\mathbf{E}_Y = -\mathbf{v} \times \mathbf{B}_Y + \eta_Y \nabla \times \mathbf{B}_Y - \alpha_Y \mathbf{B}_Y$.

Let us pass from the physical variables to the conformal ones using the conformal time $\eta = M_0/T$, $M_0 = M_{Pl}/1.66\sqrt{g^*}$, where $M_{Pl} = 1.2 \times 10^{19}$ GeV/$c^2$ is the Planck mass and $g^* = 106.75$ is the effective number of relativistic degrees of freedom.

In the FRW metric $ds^2 = a^2(\eta)(d\eta^2 - d\tilde{\mathbf{x}}^2)$ using the definitions $a = T^{-1}$, $a_0 = 1$ at temperature $T_{\text{now}}$, $d\eta = dt/a(t)$, we introduce the following notation: $\tilde{k} = ka = \text{const}$ is the conformal momentum (giving a redshift $k \sim T = T_{\text{now}}(1 + z)$); $\xi_a(\eta) = a\mu_a = \mu_a/T$ is the dimensionless fermion asymmetry that changes with time; $\tilde{\mathbf{B}}_Y = a^2\mathbf{B}_Y$ and $\tilde{\mathbf{Y}} = a\mathbf{Y}$ are the conformal dimensionless counterparts of the hypermagnetic field and hypermagnetic potential, respectively.

Here, it is convenient to rewrite (4) using the conformal coordinates $\tilde{\mathbf{x}} = \mathbf{x}/a$ for the Fourier components of the helicity density,

$$\tilde{h}_Y(\eta) \equiv \int \tilde{\mathbf{Y}} \cdot \tilde{\mathbf{B}}_Y \frac{d^3\tilde{x}}{V} = \int d\tilde{k} \tilde{h}_Y(\tilde{k}, \eta),$$

and the hypermagnetic energy density

$$\tilde{\rho}_{B_Y}(\eta) = \tilde{B}_Y^2(\eta)/2 = \int d\tilde{k} \tilde{\rho}_{B_Y}(\tilde{k}, \eta),$$

defined via their spectra

$$\begin{aligned} \tilde{h}_Y(\tilde{k}, \eta) &= \frac{\tilde{k}^2 a^3}{2\pi^2 V} \tilde{\mathbf{Y}}(\tilde{k}, \eta) \cdot \tilde{\mathbf{B}}_Y^*(\tilde{k}, \eta), \\ \tilde{\rho}_{B_Y}(\tilde{k}, \eta) &= \frac{\tilde{k}^2 a^3}{4\pi^2} \tilde{\mathbf{B}}(\tilde{k}, \eta) \cdot \tilde{\mathbf{B}}_Y^*(\tilde{k}, \eta). \end{aligned} \tag{5}$$

This allows us to calculate the integrals $\int d^3x(\ldots)/V$ in (4) in the same way as in Faraday's equation (2), by multiplying it by $\tilde{\mathbf{B}}_Y^*$ and adding with its complex conjugate $\mathbf{B}_Y \partial_t \tilde{\mathbf{B}}_Y^* = \ldots$ and to obtain the evolution equation for the helicity density spectrum and the hypermagnetic energy density spectrum.

The general system of evolution equations for the helicity density, $\tilde{h}_Y(\tilde{k}, \eta)$, and energy density, $\tilde{\rho}_{B_Y}(\tilde{k}, \eta)$, spectra satisfying the inequality $\tilde{\rho}_{B_Y}(\tilde{k}, \eta) \geq \tilde{k}\tilde{h}_Y(\tilde{k}, \eta)/2$ [15] has the following form in conformal variables:

$$\begin{aligned} \frac{d\tilde{h}_Y(\tilde{k}, \eta)}{d\eta} &= -\frac{2\tilde{k}^2}{\sigma_c} \tilde{h}_Y(\tilde{k}, \eta) \\ &+ \left(\frac{4\alpha'(\xi_{eR} + \xi_{eL}/2)}{\pi\sigma_c}\right) \tilde{\rho}_{B_Y}(\tilde{k}, \eta), \\ \frac{d\tilde{\rho}_{B_Y}(\tilde{k}, \eta)}{d\eta} &= -\frac{2\tilde{k}^2}{\sigma_c} \tilde{\rho}_{B_Y}(\tilde{k}, \eta) \\ &+ \left(\frac{\alpha'(\xi_{eR} + \xi_{eL}/2)}{\pi\sigma_c}\right) \tilde{k}^2 \tilde{h}_Y(\tilde{k}, \eta), \end{aligned} \tag{6}$$

where the constant $\alpha' = g'^2/4\pi$ is defined by the gauge SM coupling constant $g' = e/\cos\theta_W$, $\sigma_c = \sigma_{cond}/T \approx 100$ is the dimensionless plasma conductivity, $\xi_{eR}(\eta) = \mu_{eR}(T)/T$ and $\xi_{eL}(\eta) = \mu_{eL}(T)/T$ are the right- and left-handed electron asymmetries, respectively.

This system is supplemented by the kinetic equations for the asymmetries $\xi_{eR}(\eta)$ and $\xi_{eL}(\eta)$ themselves given below in Eqs. (13) and (14). It would be interesting in future to observe from Eqs. (6) how the initial field without helicity, $\tilde{h}_Y(\tilde{k}, \eta_0) = 0$, evolves in the presence of a nonzero initial energy (the initial HMF energy density spectrum) for which the derivative of the helicity density is nevertheless nonzero,

$$[d\tilde{h}_Y(\tilde{k}, \eta)/d\eta]_{\eta=\eta_0} = (4\alpha'\xi_{eR}(\eta_0)/\pi\sigma_c)\tilde{\rho}_{B_Y}(\tilde{k}, \eta_0) \neq 0.$$

The initial HMF energy density spectra, $\rho_{B_Y}(k, t_0) = Ak^{n+2}$, depend on the exponent $n$; in particular, $n = -5/3$ is substituted for the Kolmogorov spectrum. This case is the subject of a separate consideration.

For the special case of maximum helicity

$$\tilde{h}_Y(\tilde{k}, \eta) = 2\tilde{\rho}_{B_Y}(\tilde{k}, \eta)/\tilde{k} \tag{7}$$

system (6) is reduced to a single equation:

$$\frac{d\tilde{h}_Y(\tilde{k}, \eta)}{d\eta} = -\frac{2\tilde{k}^2\tilde{h}_Y(\tilde{k}, \eta)}{\sigma_c} + \left(\frac{2\alpha'[\xi_{eR}(\eta) + \xi_{eL}(\eta)/2]\tilde{k}}{\pi\sigma_c}\right)\tilde{h}_Y(\tilde{k}, \eta). \tag{8}$$

An example of such a field (a "completely helical" one), which is not considered here, that satisfies the gauge $\nabla \cdot \mathbf{Y} = 0$, $Y_0 = 0$, is the Chern–Simons wave

$$\mathbf{Y} = Y(t)(\sin k_0 z, \cos k_0 z, 0),$$

for which the HMF $\mathbf{B}_Y = \nabla \times \mathbf{Y} = k_0\mathbf{Y}$ has a nontrivial topology, being the field with maximum helicity. Indeed, its helicity density $h_Y = \mathbf{YB}_Y = k_0 Y^2(t)$ is related to the energy density $\rho_{B_Y} = \mathbf{B}_Y^2/2 = k_0^2 Y^2(t)/2$ exactly via the relation $k_0 h_Y = 2\rho_{B_Y}$.

The solution of Eq. (8) is (see also Eq. (8) in [16])

$$\tilde{h}_Y(\tilde{k}, \eta) = \tilde{h}_Y^{(0)}(\tilde{k}, \eta_0)\exp\left(\frac{2\tilde{k}}{\sigma_c}\left[\frac{\alpha'}{\pi}\int_{\eta_0}^{\eta}\left(\xi_{eR}(\eta') + \frac{\xi_{eL}(\eta')}{2}\right)d\eta' - \tilde{k}(\eta - \eta_0)\right]\right). \tag{9}$$

The spectrum of the dimensionless helicity density $\tilde{h}_Y(\tilde{k}, \eta) = a^3 h_Y(\tilde{k}, \eta)$ can be rewritten in a compact form as

$$\tilde{h}_Y(\tilde{k}, \eta) \equiv \frac{h_Y(\tilde{k}, \eta)}{T^3} = \tilde{h}_Y^{(0)}(\tilde{k}, \eta_0)\exp[A(\eta)\tilde{k} - B(\eta)\tilde{k}^2], \tag{10}$$

where the initial spectrum $\tilde{h}_Y^{(0)}(\tilde{k}, \eta_0) = h_Y(\tilde{k}, \eta_0)/T_0^3$ corresponds in our case to the instant the left-handed asymmetry appears at $T_0 = T_{RL}$. Here, we used the notation taken from (9):

$$A(\eta) = \frac{2\alpha'}{\pi\sigma_c}\int_{\eta_0}^{\eta}\left(\xi_{eR}(\eta') + \frac{\xi_{eL}(\eta')}{2}\right)d\eta', \qquad B(\eta) = \frac{2}{\sigma_c}(\eta - \eta_0). \tag{11}$$

Neglecting the quantum effects of the Abelian anomalies (in the case of $\alpha' = 0$) and in the absence of hypermagnetic diffusion (when the dynamical effects vanish in the limit of an ideal plasma, $\sigma_c \longrightarrow \infty$), from (10) we obtain the helicity density conservation law $d\tilde{h}_Y/d\eta = 0$, $\tilde{h}_Y = \text{const}$, given the conformal scaling $h_Y(\eta) = (\eta_0/\eta)^3 h_Y(\eta_0)$.

To calculate the helicity density spectrum (10), we will seek the self-consistent lepton asymmetry functions $\xi_{eR}(\eta)$ and $\xi_{eL}(\eta)$.

### *2.2. Lepton Asymmetry Evolution*

For simplicity, we consider only the inverse Higgs boson decay, i.e., neglect the Higgs boson asymmetry, $\mu_0 = 0$. The system of kinetic equations for leptons that allows for the Abelian anomalies of both right-handed and left-handed electrons (neutrinos), the inverse Higgs boson decay, and the sphaleron transitions is

$$\frac{dL_{e_R}}{dt} = \frac{g'^2}{4\pi^2 s}(\mathbf{E}_Y \cdot \mathbf{B}_Y) + 2\Gamma_{RL}\{L_{e_L} - L_{e_R}\},$$
$$\frac{dL_{e_L}}{dt} = -\frac{g'^2}{16\pi^2 s}(\mathbf{E}_Y \cdot \mathbf{B}_Y) + \Gamma_{RL}\{L_{e_R} - L_{e_L}\} - \left(\frac{\Gamma_{sph}T}{2}\right)L_{e_L}. \tag{12}$$

Here, $L_b = (n_b - n_{\bar{b}})/s \approx T^3\xi_b/6s$ is the lepton number, $b = e_R, e_L, \nu_e^L$, $s = 2\pi^2 g^* T^3/45$ is the entropy density, and $g^* = 106.75$ is the number of relativistic degree of freedom. The factor of 2 in the first row allows for the equivalence of the reaction channels $e_R\bar{e}_L \longrightarrow \varphi^{(0)}$ and $e_R\bar{\nu}_{e^L} \longrightarrow \varphi^{(-)}$; $\Gamma_{RL}$ is the decay rate (width) for Higgs bosons with the lepton chirality flip. Of course, for the left-handed doublet $L_e^T = (\nu_e^L, e_L)$, the kinetic equation for the neutrino number is redundant, because $L_{eL} = L_{\nu_{eL}}$. Further, $\Gamma_{sph} = C\alpha_W^5 = C(3.2 \times 10^{-8})$ is the dimensionless probability of the sphaleron transitions that reduce the left-handed lepton number, causing the BAU to be washed out. This probability is specified by the $SU(2)_W$ coupling constant $\alpha_W = g^2/4\pi = 1/137\sin^2\theta_W = 3.17 \times 10^{-2}$, where $g = e/\sin\theta_W$ is the gauge coupling constant in the SM and the constant $C \approx 25$ is estimated via numerical lattice calculations (see, e.g., Ch. 11 in the book [17]).

In conformal variables after integrating system (12) over the volume $\int d^3x(...)/V$ and passing to the Fourier variables for hypercharge fields, we obtain the kinetic equations (12) in the form

$$\frac{d\xi_{eR}(\eta)}{d\eta} = -\frac{3\alpha'}{\pi}\int d\tilde{k}\frac{d\tilde{h}_Y(\tilde{k},\eta)}{d\eta} - \Gamma[\xi_{eR}(\eta) - \xi_{eL}(\eta)], \tag{13}$$

$$\frac{d\xi_{eL}(\eta)}{d\eta} = \frac{3\alpha'}{4\pi}\int d\tilde{k}\frac{d\tilde{h}_Y(\tilde{k},\eta)}{d\eta} - \frac{\Gamma(\eta)}{2}[\xi_{eL}(\eta) - \xi_{eR}(\eta)] - \frac{\Gamma_{sph}}{2}\xi_{eL}(\eta), \tag{14}$$

where

$$\Gamma(\eta) = \left(\frac{242}{\eta_{EW}}\right)\left[1 - \left(\frac{\eta}{\eta_{EW}}\right)^2\right], \quad \eta_{RL} < \eta < \eta_{EW} \tag{15}$$

is the dimensionless chirality flip rate $\Gamma = 2a\Gamma_{RL}$ [9, 18], $\eta_{EW} = M_0/T_{EW} = 7 \times 10^{15}$ is the EWPT time at $T_{EW} = 100$ GeV. The derivative in the integrands of the first terms in (13) and (14), $d\tilde{h}_Y(\tilde{k}, \eta)/d\eta$, is given by Eq. (8), where we should substitute $\tilde{h}_Y(\tilde{k}, \eta)$ from Eq. (10) on the right-hand side.

We choose the following initial conditions at the time $\eta_0 = \eta_{RL} = 7 \times 10^{13}$, corresponding to the temperature $T_{RL} = 10$ TeV:

$$\xi_{eL}(\eta_0) = 0, \quad \xi_{eR}(\eta_0) = 10^{-10}. \tag{16}$$

In Section 3.1, we also discuss the case of a large initial lepton asymmetry, $\xi_{eR}(\eta_0) = 10^{-4}$, because this is a free parameter in our problem.

The solution of system (13) and (14) allows the evolution of the hypermagnetic helicity density (10) to be calculated for two cases:

(i) a monochromatic helicity density spectrum

$$\tilde{h}_Y(\tilde{k}, \eta) = \tilde{h}_Y(\eta)\delta(\tilde{k} - \tilde{k}_0), \tag{17}$$

(ii) a continuous initial spectrum $\tilde{h}_Y(\tilde{k}, \eta_0) \sim \tilde{k}^{n_s}$, $n_s \geq 3$.

Here, the initial helicity density for the monochromatic spectrum (17) $\tilde{h}_Y(\eta_0) = (\tilde{B}_0^Y)^2/\tilde{k}_0$ is given by the seed field $\tilde{B}_0^Y$. The problem has two free parameters: (i) the seed field $\tilde{B}_0^Y$ at the initial temperature $T_0 = T_{RL} = 10$ TeV and (ii) the initial right-handed electron asymmetry $\xi_{eR}(\eta_0) \neq 0$ in the chosen scenario [9, 10]. We everywhere assume the initial hypermagnetic energy density to be $\tilde{\rho}_{B_Y}^{(0)} = 10^{-8}$, corresponding to a strong seed field $B_0^Y = 10^{-4}\sqrt{2}T_0^2 \sim 10^{24}$ G. Note that such a field does not affect the Friedmann law of expansion of the Universe, $\rho_{B_Y} \ll \rho_\gamma \sim T^4$.

## 3. CONSERVATION LAWS AND THE BAU IN HYPERMAGNETIC FIELDS

As follows from the kinetic equations (12), in the absence of a hypercharge field, the total lepton number is not conserved due to the sphaleron transitions that wipe out the left-handed leptons, $dL_e/dt = \dot{L}_{e_R} + \dot{L}_{e_L} + \dot{L}_{\nu_{e}^{L}} = -\Gamma_{sph}L_{e_L}$. Baryogenesis is realized via leptogenesis due to the conservation law $B/3 - L_e = \text{const}$, where $B = (n_B - n_{\bar{B}})/s$. Given the Abelian anomalies in system (12), such baryogenesis is possible, $\dot{B} \neq 0$, when the HMF increases the lepton number and the BAU, $dL_e/dt|_{B_Y \neq 0} > 0$ and $dB/dt|_{B_Y \neq 0} > 0$. This process competes with the influence of sphalerons washing out $L_{eL}$ and $B$ (for comparison, see [9] where we neglected the sphaleron transitions).

Three global charges are conserved ($\delta_i = \text{const}$):

$$\frac{B}{3} - L_e = \delta_1, \quad \frac{B}{3} - L_\mu = \delta_2, \quad \frac{B}{3} - L_\tau = \delta_3, \tag{18}$$

as well as $L_{e_R} = \delta_R$ as long as $T \gg T_{RL}$. If the initial BAU differs from zero, $B(t_0) \neq 0$, and if we assume the absence of a lepton asymmetry for the second- and third-generation particles up to $T_{EW}$, $L_\mu = L_\tau = 0$, then we find that the relation $\delta_2 = \delta_3 = B(x_0)/3$ is valid only at the initial time. We find the change in BAU, $B(t)$, at temperatures $T < T_{RL}$ from the first conservation law in Eq. (18). This change obeys the relation

$$\frac{B(t)}{3} - L_e(t) = \frac{B(t_0)}{3} - L_{e_R}(t_0) = \delta_{2,3} - \delta_R = \delta_1.$$

If, for simplicity, we assume the initial BAU to be zero, $B(t_0) = 0$, or $\delta_{2,3} = 0$, then, as a result, we obtain the conservation law $B(t)/3 - L_e(t) = -L_{e_R}(t_0)$.

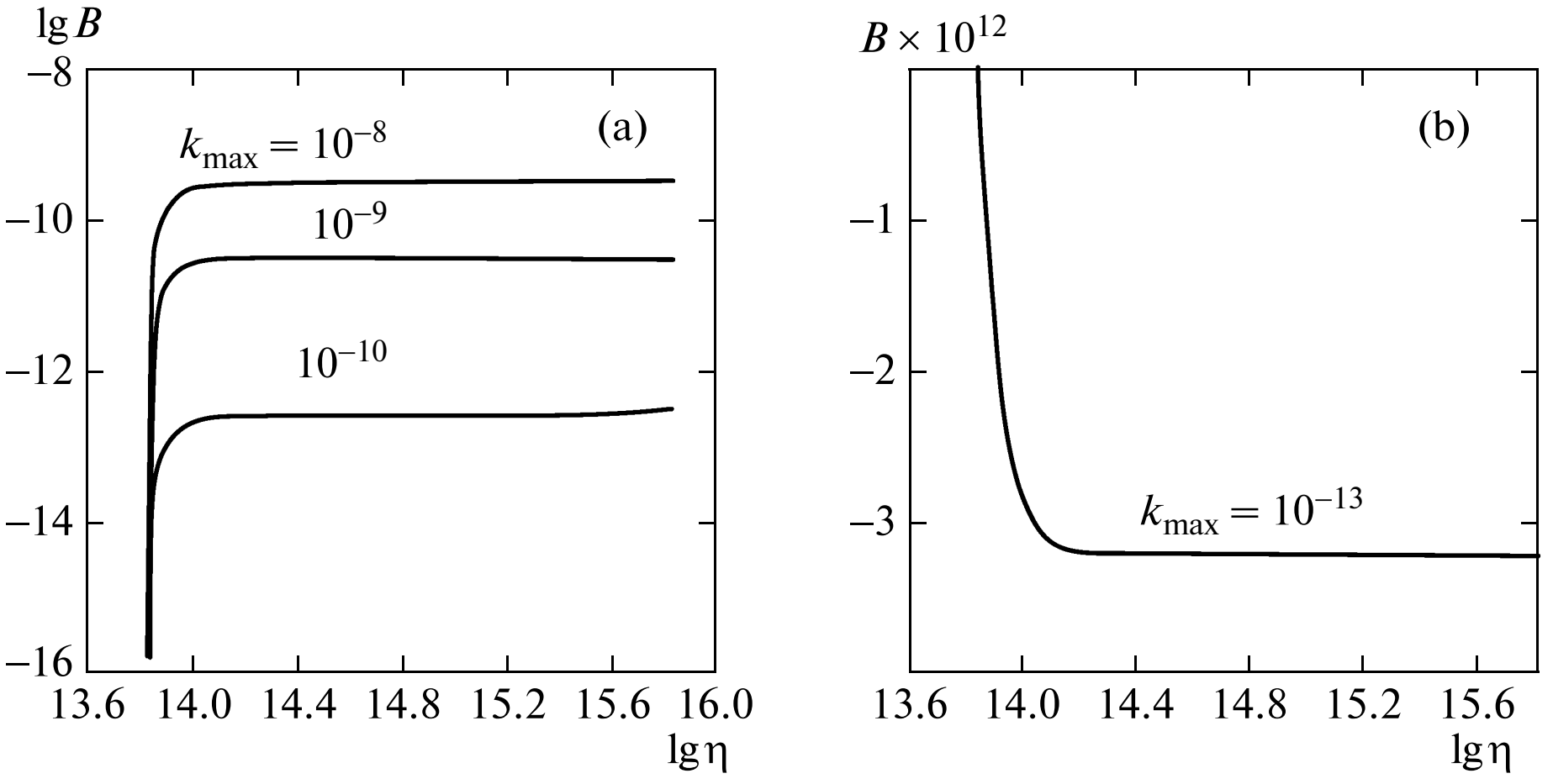

**Fig. 1.** (a) BAU evolution $B(\eta)$ on a logarithmic scale for a continuous initial helicity density spectrum $\tilde{h}_Y(\eta_0, \tilde{k}) = C\tilde{k}^{n_s}$, $n_s =$ 3, and initial right-handed asymmetry $\xi_{eR}(\eta_0) = 10^{-10}$. (b) Negative BAU $B(\eta) < 0$ evolving for the same spectrum and initial asymmetry $\xi_{eR}(\eta_0) = 10^{-10}$ for the minimum possible wave number $\tilde{k}_{\max} = 10^{-13}$. In both cases, the initial left-handed asymmetry is absent: $\xi_{eL}(\eta_0) = 0$.

Thus, in this case, the BAU "sits" in a hypercharge field and decreases due to the sphaleron transitions, as follows from the kinetic equations (12):

$$B(t) = 3\int_{t_0}^{t}\left[\frac{dL_{e_R}(t')}{dt'} + \frac{dL_{e_L}(t')}{dt'} + \frac{dL_{\nu_e^L}(t')}{dt'}\right]dt' = \frac{3g'^2}{8\pi^2}\int_{t_0}^{t}(\mathbf{E}_Y \cdot \mathbf{B}_Y)\frac{dt}{s} - 3\int_{t_0}^{t}\Gamma_{sph}TL_{e_L}dt'. \quad (19)$$

Using the first equation in system (12), where the hypermagnetic term originates from the Abelian anomaly $\sim(\mathbf{E}_Y \cdot \mathbf{B}_Y)$, we obtain the baryon asymmetry in the following form from Eq. (19):

$$B(\eta) = 5.3 \times 10^{-3}\int_{\eta_0}^{\eta}d\eta'\left\{\frac{d\xi_{e_R}(\eta')}{d\eta} + \Gamma(\eta) \times [\xi_{e_R}(\eta') - \xi_{e_L}(\eta')]\right\} - \frac{6 \times 10^7}{\eta_{EW}}\int_{\eta_0}^{\eta}\xi_{e_L}(\eta')d\eta'. \quad (20)$$

### *3.1. BAU Evolution for a Continuous Helicity Density Spectrum*

The baryon asymmetry evolution in HMFs with maximum helicity $\tilde{k}\tilde{h}_Y(\eta, \tilde{k}) = 2\tilde{\rho}_{B_Y}(\eta, \tilde{k})$ is described by Eq. (20) and is shown in Fig. 1.

The hypermagnetic helicity density spectrum $\tilde{h}_Y(\tilde{k}, \eta)$ plays a crucial role for the BAU evolution, as follows from the above kinetic equations (13) and (14). For a continuous initial spectrum

$$\tilde{h}_Y(\tilde{k}, \eta_0) = C_1\tilde{k}^{n_s} \quad (21)$$

we define the helicity density as

$$\tilde{h}_Y(\eta) = C_1\int_0^{\tilde{k}_{\max}}\tilde{k}^{n_s}\exp[A(\eta)\tilde{k} - B(\eta)\tilde{k}^2]d\tilde{k} = C_1 I_{n_s}(\eta). \quad (22)$$

Here, the functions $A(\eta)$ and $B(\eta)$ are given by Eq. (11). The constant $C_1$ can be estimated using the relation for a completely helical field

$$\tilde{h}_Y(\tilde{k}, \eta_0) = C_1\tilde{k}^{n_s} = 2\tilde{\rho}_{B_Y}(\tilde{k}, \eta_0)/\tilde{k}.$$

Using the definition of the initial hypermagnetic energy

$$\int d\tilde{k}\tilde{\rho}_{B_Y}(\tilde{k}, \eta_0) = (\tilde{B}_0^Y)^2/2,$$

we obtain the relation

$$C_1\int_0^{\tilde{k}_{\max}}\tilde{k}^{n_s+1}d\tilde{k} = (\tilde{B}_0^Y)^2 = 2\tilde{\rho}_Y^{(0)} = 2 \times 10^{-8}$$

for the seed field chosen above. Then, we vary the maximum value $k_{\max}$ proportional to the hypermagnetic diffusion efficiency: the shorter the wavelength, the stronger the HMF diffusion. Thus, we determine the constant $C_1 = (n_s + 2)(\tilde{B}_0^Y)^2/(k_{\max})^{n_s+2}$.

In the case of a continuous initial spectrum (21), we can rewrite the kinetic equations for the lepton

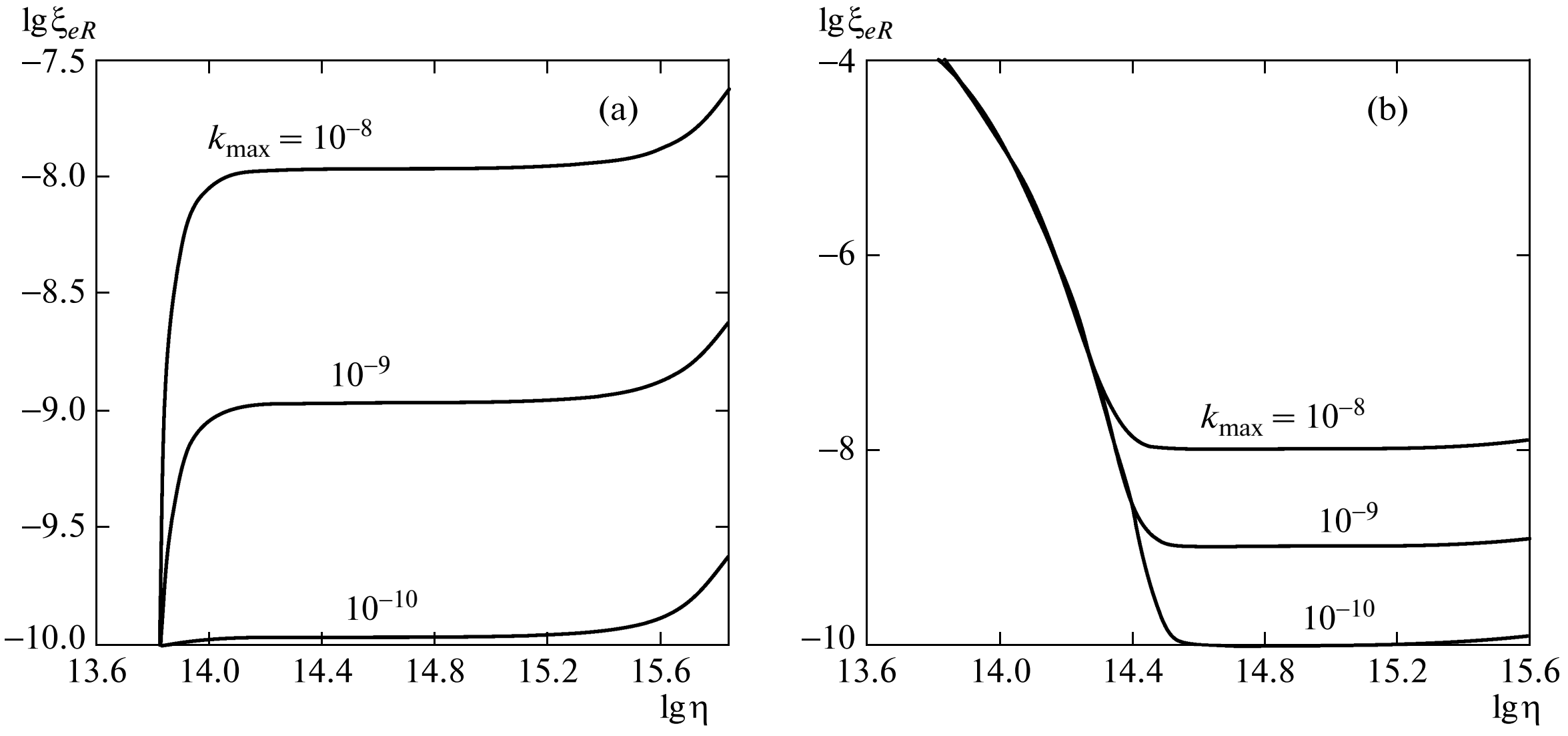

**Fig. 2.** Leptogenesis $\xi_{eR}(\eta)$ on a logarithmic scale for a continuous initial helicity density spectrum $\tilde{h}_Y(\eta_0, \tilde{k}) = C\tilde{k}^{n_s}$, $n_s = 3$, $0 \leq \tilde{k} \leq \tilde{k}_{max}$: (a) three curves for various wave numbers $\tilde{k}_{max} = 10^{-8}, 10^{-9}$, and $10^{-10}$ start from the initial value of $\xi_{eR}(\eta_0) = 10^{-10}$; (b) for the same values of $\tilde{k}_{max}$, the curves start from $\xi_{eR}(\eta_0) = 10^{-4}$. The initial left-handed lepton asymmetry is zero in both cases, $\xi_{eL}(\eta_0) = 0$.

asymmetry (13) and (14) governing the change in BAU as

$$\frac{d\xi_{eR}}{d\eta} = \frac{6\alpha' C_1}{\pi\sigma_c}\left[I_{n_s+2}(\eta) - \frac{\alpha'}{\pi}\left(\xi_{eR} + \frac{\xi_{eL}}{2}\right)I_{n_s+1}(\eta)\right] - \Gamma(\eta)(\xi_{eR} - \xi_{eL}), \quad (23)$$

$$\frac{d\xi_{eL}}{d\eta} = -\frac{3\alpha' C_1}{2\pi\sigma_c}\left[I_{n_s+2}(\eta) - \frac{\alpha'}{\pi}\left(\xi_{eR} + \frac{\xi_{eL}}{2}\right)I_{n_s+1}(\eta)\right] - \Gamma(\eta)(\xi_{eL} - \xi_{eR}) - \frac{\Gamma_{sph}}{2}\xi_{eL}(\eta). \quad (24)$$

The integrals $I_{(n_s+2),(n_s+1)}(\eta)$ are functions of the lepton asymmetries $\xi_{eR}$ and $\xi_{eL}$ via $A(\eta)$ in Eq. (22); thus, these differential equations are highly nonlinear and can be solved only numerically.

Figure 2 shows the evolution of the right-handed lepton asymmetry $\xi_{eR}(\eta)$ found by solving the system of self-consistent equations (23) and (24). This can help us to interpret the BAU evolution in Figs. 1 and 3. Note that the left-handed lepton asymmetry $\xi_{eL}$ is much smaller, $\xi_{eL} \ll \xi_{eR}$, first, due to the sphaleron transitions reducing $L_{eL}$ and, second, due to the initial conditions $\xi_{eL}(\eta_0) = 0$ and $\xi_{eR}(\eta_0) \neq 0$ under which $\xi_{eL}$ has no time to grow by the EWPT time $\eta_{EW}$. Indeed, assuming that $\partial_t\xi_{eR} = \partial_t\xi_{eL} \approx 0$ at the saturation level, multiplying (14) by 4, and adding (13), we obtain

$$\xi_{eL} = \frac{\Gamma\xi_{eR}}{\Gamma + 2\Gamma_{sph}} \ll \xi_{eR}, \quad (25)$$

where $\Gamma_{sph} \gg \Gamma$. The explanation of why $\xi_{eR}(\eta)$ grows due to the Abelian anomaly (1), tending asymptotically to the saturation level, $\xi_{eR}(\eta) \approx \text{const}$, is given in [8], where such a saturation level is shown to be independent of the chosen initial condition $\xi_{eR}(\eta_0) = 10^{-10}$ or $\xi_{eR}(\eta_0) = 10^{-4}$ in the case of a monochromatic helicity density spectrum. Similarly, the absence of any dependence of the saturation level $\xi_{eR} \approx$ const also becomes obvious here for a continuous helicity density spectrum when Figs. 2a and 2b are compared for identical $\tilde{k}_{\max}$. Then, the additional growth of the right-handed lepton asymmetry by the end of the interval in Fig. 2 is explained by the disappearance of the inverse Higgs decay (15) and leads to an additional (second) BAU growth in Fig. 1 as $\eta \longrightarrow \eta_{EW}$.

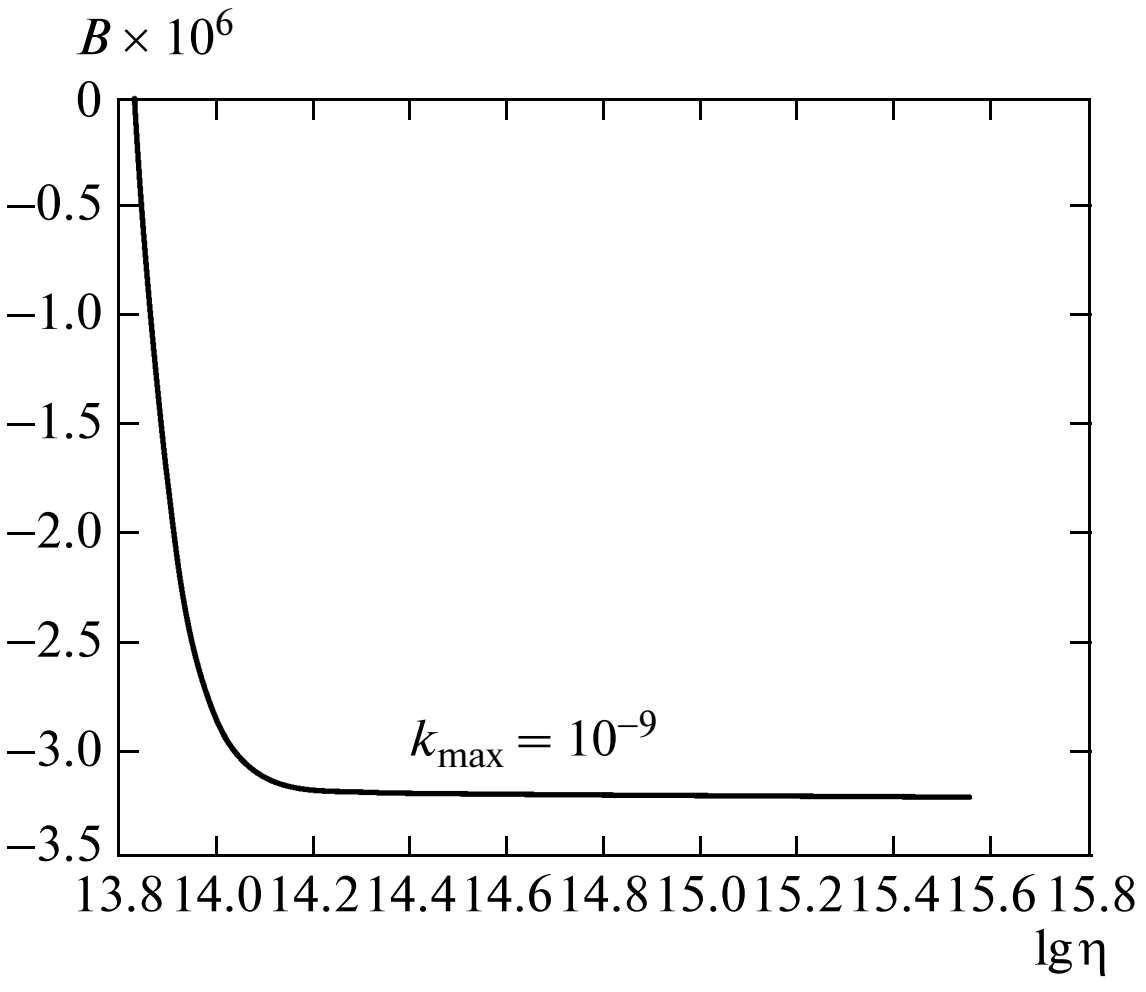

**Fig. 3.** Negative baryon asymmetry for a continuous initial helicity density spectrum $\tilde{h}_Y(\eta_0, \tilde{k}) = C\tilde{k}^{n_s}$, $n_s = 3$. The BAU line corresponds to $\tilde{k}_{max} = 10^{-9}$, where the right-handed asymmetry $\xi_{eR}(\eta)$ starts from a large initial value, $\xi_{eR}(\eta_0) = 10^{-4}$. The initial left-handed asymmetry is zero, $\xi_{eL}(\eta_0) = 0$.

## DISCUSSION

We considered the leptogenesis and the corresponding baryogenesis in the presence of hypermagnetic fields before the EWPT time, $T > T_{EW} \approx 100$ GeV, when the Abelian anomaly for $e_R$ and the left-handed doublet $L = (\nu_e^L e_L)^T$ provides the evolution of their asymmetries. We took into account the inverse Higgs decay and the sphaleron transitions in a wide temperature range, $T_{RL} \geq T > T_{EW}$, $T_{RL} \approx 10$ TeV. The doubts as to whether it is possible to provide the observed baryon asymmetry of the Universe in its symmetric phase by temporarily "storing" the BAU in the $e_R$ asymmetry are dispelled in the case of strong HMFs. The BAU washout by the sphaleron transitions due to the involvement of left-handed particles $T < T_{RL}$ via the inverse Higgs decay is not critical in a wide range of HMF strengths. A strong seed HMF $B_Y^{(0)}$ guarantees the needed BAU growth.

BAU growth is possible only in the case of a growing right-handed electron asymmetry, $d\xi_{eR}(\eta)/d\eta > 0$, starting from a small initial value, $\xi_{eR}(\eta_0) = 10^{-10}$ (see Fig. 2a). However, even for positive $d\xi_{eR}(\eta)/d\eta > 0$, such growth is not possible for all HMF scales $\Lambda = \tilde{k}^{-1}$ in the range of wave numbers $0 < \tilde{k} \leq \tilde{k}_{\max}$. The smaller $k_{\max}$, the smaller the BAU growth due to the decrease in helicity density $h_Y \approx YB \sim kY^2$ as a source of leptogenesis via the Abelian anomaly. As a result, the BAU growth ceases for small $k_{\max}$ and instead we see a decrease in BAU down to negative values, $B < 0$. Note that a similar dependence was found for a monochromatic Chern–Simons wave (see the right panel of Fig. 1 in [10]). In this paper, Fig. 1b shows a drop in BAU that becomes negative, $B < 0$, almost immediately for small $\tilde{k}_{\max} = 10^{-13}$.

On the other hand, such a free parameter as a large initial lepton asymmetry ($\xi_{eR}(\eta_0) = 10^{-4}$) does not allow a positive BAU, $B > 0$, to be obtained. Indeed, despite the same saturation level for $\xi_{eR}$ for both initial conditions, small $\xi_{eR}(\eta_0) = 10^{-10}$ and large $\xi_{eR}(\eta_0) = 10^{-4}$ (see Fig. 2), the negative sign of the derivative $d\xi_{eR}/d\eta < 0$ in the second case (see also Eq. (20)) leads to a negative BAU, $B < 0$. This case is shown in Fig. 3 for a large initial asymmetry, $\xi_{eR}(\eta_0) = 10^{-4}$, and the range of wave numbers $0 \leq \tilde{k} \leq \tilde{k}_{\max} = 10^{-9}$, for which, on the contrary, BAU growth was observed in the case of a small initial asymmetry, $\xi_{eR}(\eta_0) = 10^{-10}$ (for comparison, see Fig. 1a).

We emphasize the difference between the monochromatic and more realistic continuous helicity density spectra in their influence on the BAU growth. The case of monochromatic and continuous magnetic helicity density spectra has recently been considered in [8] without calculating the corresponding BAU. Nevertheless, such BAU evolution is shown in Fig. 1 in [10], where the Chern–Simons wave of a hypercharge field $Y_\mu$ with some fixed wave numbers $\tilde{k}_0$ was considered. Note that the Chern–Simons wave has a maximum helicity density (see the comments to Eq. (9) above), making the comparison with the case of a continuous spectrum reasonable. The solid line in Fig. 1 from [10] reaches $B_{\text{obs}} \sim 10^{-10}$ for $\tilde{k}_0 \sim 10^{-10}$ using the parameter $B_0 = 2.1 \times 10^{-2}$ in Eq. (3.8), while in this paper the case of $k_{\max} \sim 10^{-10}$ leads to a small BAU, $B \ll B_{\text{obs}}$, and only a large $\tilde{k}_{\max} \sim 10^{-8}$–$10^{-9}$ allows $B = B_{\text{obs}} \sim 10^{-10}$ to be obtained (see Fig. 1a). Such a contrasting difference is explained by allowance for the inverse cascade that reduces the wave numbers $\tilde{k} < \tilde{k}_{\max}$, $\tilde{k} \longrightarrow 0$, and the large-scale HMFs themselves $B_Y \sim kY$, which have progressively smaller amplitudes in the successive steps of the inverse cascade, and also by the fact that the helicity density $h_Y \sim YB_Y \sim kY^2$ drops, ceasing to maintain the growth of the lepton number and the BAU growth.

It can be concluded that an observed baryon asymmetry $B_{\text{obs}} \sim 10^{-10}$ can be provided by leptogenesis in a strong HMF for a wide range of HMF scales $\Lambda = k^{-1}$ specified by the wave numbers $0 < \tilde{k} \leq \tilde{k}_{\max}$ in the continuous spectrum. For a small initial lepton asymmetry, $\xi_{eR}(\eta_0) = 10^{-10}$, $B_{\text{obs}} \sim 10^{-10}$ can be obtained asymptotically for a spectrum limited by the interval $\tilde{k}_{\max} \sim 10^{-8}$–$10^{-9}$. For the same initial asymmetry, a narrower inverse cascade in the range of wave numbers $0 < \tilde{k} < \tilde{k}_{\max} < 10^{-10}$ leads to a smaller BAU, $B \ll B_{\text{obs}}$ (see Fig. 1a). This occurs due to the reduced helicity density as a BAU source when integrating in a narrower range of the continuous spectrum.

## ACKNOWLEDGMENTS

We thank D.D. Sokoloff for the discussion and useful remarks.